\documentclass[prb,amsmath,amsfonts,showpacs]{revtex4}
\usepackage{graphicx}
\usepackage{epsf}

\begin{document}
\title{Charge fluctuations and feedback effect in shot noise in a Y-terminal system}
\author{Bogdan R. Bu{\l}ka}
\affiliation{Institute of Molecular Physics, Polish Academy of Sciences,
ul.~M.~Smoluchowskiego~17, 60-179~Pozna\'{n}, Poland}

\date{Received \today \hspace{5mm} }

\begin{abstract}
We investigate a dynamical Coulomb blockade effect and its role in the
enhancement of current-current correlations in a three-terminal device with a
multilevel splitter, as well as with two quantum dots. Spectral decomposition
analysis shows that in the Y-terminal system with a two level ideal splitter,
charge fluctuations at a level with a lowest outgoing tunneling rate are
responsible for a super-Poissonian shot noise and positive cross-correlations.
Interestingly, for larger source-drain voltages, electrons are transferred as
independent particles, when three levels participate in transport, and double
occupancy is allowed. We can explain compensation of the current correlations
as the interplay between different bunching and antibunching processes by
performing a spectral decomposition of the correlation functions for partial
currents flowing through various levels. In the system with two quantum dots
acting as a splitter, a long range feedback effect of fluctuating potentials
leads to the dynamical Coulomb blockade and an enhancement of shot noise.

\end{abstract}

\maketitle

\section{Introduction}

For a long time, the interest in current shot noise in nanoscale
systems was mostly focused on its reduction from the Schottky's value
$S_{P}=2eI$, where $I$ is the average current, and $e$ is the charge
of an electron.  The shot noise value $S_{II}$ is reduced because of
the Pauli exclusion principle of scattered particles, and for the
ballistic point contacts, it can be reduced to
zero.~\cite{kulik,khlus,lesovik,buttiker1990} In metallic diffusive
wires, the reduction reaches $1/3\times 2eI$,~\cite{beenakker1992} and
for chaotic cavities, in the classic limit, $1/4\times
2eI$~\cite{baranger,jalabert}, whereas for sequential tunneling through a
quantum dot (QD), a maximal reduction is $1/2\times 2eI$
~\cite{korotkov92,hershfield,korotkov} (for an extensive review see
[\onlinecite{blanter}]).

Electronic correlations can lead to an increase of shot noise above
 the Schottky's value (so called super-Poissonian shot
 noise). Iannoccone at al.~\cite{iannaccone} showed that shot noise in
 a resonant-tunneling diode, biased in the negative differential
 resistance regions of the $I$-$V$ characteristic, is enhanced because
 of the raise of the well's potential energy, which causes more states
 to be available for successive tunneling events from the
 cathode. Kuznetsov et al.\cite{kuznetsov} interpreted a transition of
 shot noise from the sub-Poissonian to super-Poissonian regime in the
 quantum well as a result of a change of the shape of the density of
 states, in which a parallel magnetic field leads to multiple voltage
 ranges of negative differential resistance. Our studies of a
 ferromagnetic single electron transistor showed~\cite{bb99} an
 increase of $S_{II}$, much above $S_P$ at the pinch-off voltage; that
 is, when the current begins to flow, and strong back-scattering leads
 to an enhancement in shot noise (see also
 [\onlinecite{safonov}]). Asymmetry in conducting channels for
 electrons with opposite spins can activate spin fluctuations, which
 results in the super-Poissonian current shot noise.~\cite{bb99,bb00}
 We call the effect the dynamical Coulomb blockade because the
 electron that cannot leave for some time the QD blocks the channel
 for an electron with the opposite spin.  Experimental efforts have
 just recently been undertaken~\cite{aliev,garzon}, in order to search
 for super-Poissonian shot noise in magnetic tunnel-junctions. The
 dynamical Coulomb blockade effect can also be seen in a nonmagnetic
 system of two capacitively coupled quantum dots connected in parallel
 to external electrodes.~\cite{michalek} In this case, fluctuations of
 charge polarization can result in an increase of current shot
 noise. Recent measurements were performed on such devices by McClure at
 al.~\cite{mcclure} and Zhang et al.~\cite{zhang} for current noise
 auto-correlations and cross correlations, and confirmed in part the theoretical
 predictions. Sukhorukov et al.\cite{sukhorukov} and Thielmann et
 al.\cite{thielmann} suggested that inelastic spin-flip cotunneling
 processes can also lead to super-Poissonian shot noise, which can be
 observable for bias voltages around the corresponding energy for
 spin-flip excitations.

In a multi-terminal geometry, one can study the statistics of scattered
particles in cross correlation functions.\cite{feynman} It is well known as the
Hanbury Brown and Twiss experiment (HBT)\cite{hbt}, involving two incoming
particle streams and two detectors showing bunching for photons. The HBT
experiments for electrons were performed on a quantum point
contact\cite{liu,oliver} and with fractional quantum Hall effect (FQHE) edge
states\cite{henny}, and they showed an antibunching effect because of the Pauli
principle. The cross-correlation in outgoing channels is FQHE effect negative
in such cases\cite{loudon,texier} (see also [\onlinecite{martin}] for cross
correlations in a Y-terminal geometry). Texier and B{\"u}ttiker
showed~\cite{texier} that inelastic scattering can lead to positive
current-current correlations in a multiterminal system with FQHE edge states,
whereas correlations remain always negative for quasielastic scattering. In
their model, an additional electrode was introduced to keep the current equal
to zero, and caused voltage fluctuations. Recently, Oberholzer et
al.\cite{oberholzer} verified experimentally these predictions in a device, in
which interactions between current carrying states and fluctuating voltage were
controlled by an external gate voltage. Using general arguments for scattering
of quantum particles~\cite{feynman}, Burkard et al.\cite{burkard} showed that
an entangled singlet electron pair gives rise to an enhancement of the noise
power (bunching), whereas the triplet pair leads to a suppression of noise
(antibunching). Positive cross-correlations were predicted by Cottet et
al.\cite{cottetprb,cottetprl,cottetepl} in sequential tunneling through a
three-terminal QD system. The dynamical Coulomb blockade
effect\cite{bb99,bb00,michalek} between electrons transferred through a
multilevel QD can lead to the super-Poissonian effect for current-current
auto-correlation functions, as well as to positive cross-correlation
functions.~\cite{cottetprb,cottetprl,cottetepl} They also analyzed time
evolution of tunneling events and presented the bunching effect in the system.
Gustavsson et al.\cite{gustavssonprl,gustavssonprb} recently took time-resolved
measurements of electron transport through a multilevel QD, using a nearby
quantum point contact as a charge detector. They were able to detect bunching
of electrons, leading to the super-Poissonian shot noise. A phenomenological
approach to current cross correlations was presented by Wu and Yip\cite{wu}.
They used a Langevin formalism in circuit modeling taking into account voltage
and current fluctuations. This general method was applied for calculation of
current cross-correlation functions in a Y-terminal shaped system. In a similar
manner, Rychkov and B{\"u}ttiker\cite{rychkov} showed that current cross
correlations are always positive in a macroscopic classical Y-terminal system
with a fluctuating current in an input branch.

In this paper, we would like to present a study of a microscopic nature of shot
noise in a multiterminal geometry. Our motivation are the recent experimental
studies~\cite{chen,zarchin} of shot noise on a quantum point contact, acting as
a beam splitter, and those~\cite{zhang} in two capacitatively coupled quantum
dots. In the quantum point contact system, the measurements showed an
enhancement of shot noise much above the Schottky value\cite{zarchin} and
positive current-current cross correlations\cite{chen}. Unfortunately, these
experiments~\cite{chen,zarchin} do not give an answer to the origin of bunching
of electrons. In the two quantum dot system, the shot noise results are
interpreted~\cite{zhang} within a model for a single quantum dot with a
multilevel structure. The interpretation ignores spatial dynamical charge
fluctuations, which occur in tunneling through the coupled quantum dots and
their influence of shot noise. We expect that the dynamical Coulomb blockade
effect is responsible for bunching electrons in both the systems, but its
origin is different. In a multi-level quantum dot, local charge fluctuations
are relevant. On the other hand, the potential feedback, caused by spatial
charge fluctuations, leads to bunching in a multi-dot system. Therefore, our
goal is to study quantitatively current characteristics and current correlation
functions in sequential transport in a three-terminal model with a multi-level
splitter and with two-quantum dots.

In the first part of the paper, we will consider the Y-terminal system with a
two-level splitter showing that super-Poissonian shot noise and positive
cross-correlations can be expected for large asymmetry in outgoing tunneling
rates, when a dynamical Coulomb blockade effect can occur. One could naively
expect an enhancement of shot noise with an increase of a number of levels in
the splitter. However, the shot noise is reduced when the third level in the
splitter becomes to participate in transport. We will show an interplay of auto
and cross-correlation partial contributions to a total shot noise and its
reduction in the Y-terminal system with a three-level splitter. In a
multi-terminal system, one can change the number of levels participating in
transport to different electrodes by applying different voltage potentials. We
will also present how the current auto and cross-correlation functions can
evolve when a voltage window is changed in one of the electrode.

The second part of the paper is devoted to the Y-terminal system with two
quantum dots. Our considerations will be focused on a feedback effect of
potential fluctuations at quantum dots and its role in enhancement of shot
noise in the current-current auto- and cross-correlation functions. We will
show that an origin of electron bunching is in this case similar, and it is
caused by charge accumulation at one QD. Consequently, this results in strong
dynamical Coulomb blockade effect. When a second electron participates in
transport (for higher bias voltages), antibunching processes dominate, and the
correlation functions show features typical for sub-Poissonian shot noise. Our
analysis (in the Appendix) will show that shot noise reduction can be
substantial, and the Fano factor can reach its minimal value $F=1/(N+1)$, in a
system of $N$ quantum dots connected in series.

\section{Y-terminal system with a two level splitter}

\subsection{General derivations}

\begin{figure}[ht]
\centerline{\epsfxsize=0.35\textwidth \epsfbox{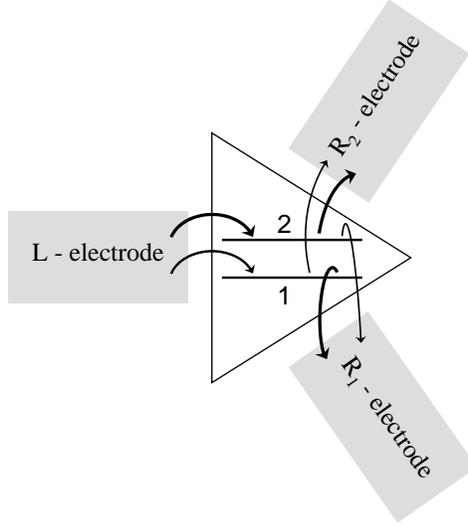} }
\caption{\label{fig1} Schematic overview of the Y-terminal system with two
level splitter, which transfers electrons from the source electrode to the two
drain electrodes (from the left to the right hand side).}
\end{figure}

Let us consider a three-terminal system with a two-level QD as a splitter
(presented schematically in Fig.\ref{fig1}).
The electronic transport in a
sequential regime is governed by the classical master equation~\cite{korotkov,gurvitz96,gurvitz98}
\begin{eqnarray}\label{b1}
\frac{d}{dt} \left[\begin{array}{c}
p_3\\p_2\\p_1\\p_0\end{array}\right]=\hat{M}\left[\begin{array}{c}
p_3\\p_2\\p_1\\p_0\end{array}\right]\;,
\end{eqnarray}
where $p_0$, $p_1$, $p_2$ and $p_3$ denote probabilities of finding an empty
system, a system with one electron at the level $E_1$ or at the level $E_2$, and a
probability of a system with two electrons,
respectively. The matrix $\hat{M}$ is given by
\begin{eqnarray}\label{b2}
\hat{M}=\left[\begin{array}{cccc}
-\Gamma^{31}-\Gamma^{32}&\Gamma^{23}&\Gamma^{13}&0\\
\Gamma^{32}&-\Gamma^{20}-\Gamma^{23}&0&\Gamma^{02}\\
\Gamma^{31}&0&-\Gamma^{10}-\Gamma^{13}&\Gamma^{01}\\
0&\Gamma^{20}&\Gamma^{10}&-\Gamma^{01}-\Gamma^{02}\\
\end{array}\right]\;,
\end{eqnarray}
where $\Gamma^{nm}$ is the tunneling rate, which changes the occupation of the
splitter from the state $n$ to $m$,
$\Gamma^{nm}=\sum_{\alpha}\Gamma^{nm}_{\alpha}$, and $\Gamma^{nm}_{\alpha}
=f_{\alpha,n}^{\pm}\gamma^{nm}_{\alpha}$ denotes the total tunneling rate from
the $\alpha$ electrode ($\alpha=L$, $R_1$, $R_2$). $\gamma^{nm}_{\alpha}$ is
the net tunneling rate through the $\alpha$ tunnel junction,
$f^{\pm}_{\alpha,n}=1/\{1+\exp[\pm(E_{n}-\mu_{\alpha})/k_BT]\}$, the sign $+$
(or $-$) is for a transfer an electron from (or to) the electrode, and
$\mu_{\alpha}$ denotes the chemical potential in the ${\alpha}$ electrode. We
assume that the energy levels $E_1<E_2$, and when two electrons occupy the
splitter, the corresponding energy is $E_3=E_1+E_2+U_{12}$, which includes the
electron-electron repulsion energy $U_{12}$.

The current from the $\alpha$-th electrode is given by
\begin{eqnarray}\label{b3}
I_{\alpha}\equiv I_{\alpha}^+-I_{\alpha}^-= e\sum_{n<m}(\Gamma^{nm}_{\alpha}
p_{m}-\Gamma^{mn}_{\alpha} p_{n})\;.
\end{eqnarray}
(The electronic charge in our notation is taken as $e$.) The probabilities
$p_n$ at the stationary state are determined from the master equation (\ref{b1})-(\ref{b2})
with the left hand side equal to zero.

In the sequential regime, it is assumed that tunneling events are independent,
corresponding tunneling resistances $R^{nm}_{\alpha}\gg R_Q = h/2e^2$, and the
electronic transport is dominated by sequential tunneling
processes~\cite{averin,schon}, whereas higher order processes (cotunneling) are
neglected. Moreover, it is assumed that the resonant current peak for each
energy level (described by the Breit-Wigner formula) is strongly broadened by
temperature; i.e., $h\gamma^{nm}_{\alpha}\ll k_BT$. Comparing our results on the
shot noise with an experiment, one has to extract a thermal noise contribution,
which is always present in any conductor (e.g., see [\onlinecite{dicarlo}] for
thermal calibration procedure of an experimental setup).

 Fluctuations in the system are
studied within the generation-recombination approach for multi-electron
channels.~\cite{vliet,korotkov} The Fourier transform of the correlation
function of the quantity $X$ and $Y$ can be expressed as~\cite{korotkov}
\begin{eqnarray}\label{b4}
S_{XY}(\omega)\equiv 2\int_{-\infty}^{\infty}dt e^{i\omega t}\left[\langle
X(t)Y(0)\rangle-\langle X \rangle\langle Y\rangle\right]
=4\sum_{n,m=0,1,2,3}X_{n}\left[P_{nm}(\omega)-\frac{p_n}{i\omega}\right]Y_mp_m\;,
\end{eqnarray}
where $p_n$ is the stationary value, $X_n$, and $Y_n$ are the values of $X$ and
$Y$ at this state. The conditional probability $P(n,m;t)$ to find the system in
the state $n$ at time $t$, if it was in the initial state $m$ at $t=0$,
satisfies the master equation (\ref{b1}),~\cite{vliet,korotkov} and its Fourier
transform is given by $P(n,m;\omega)=[i\omega\hat{1}-\hat{M}]^{-1}_{nm}$. The elements
of the Green's function
\begin{eqnarray}\label{b4a}
G(n,m;\omega)\equiv
[i\omega\hat{1}-\hat{M}]^{-1}_{nm}-p_n/i\omega
\end{eqnarray}
 can be determined directly by  matrix inversion. It is useful to perform
 spectral decomposition, and analyze fluctuations corresponding to
 characteristic eigenfrequencies. Therefore, the Green function (\ref{b4a}) is represented
 by its eigenvalues  $\lambda_\nu$ and the corresponding matrix of
 eigenvectors $\hat{R}$ as
\begin{eqnarray}\label{b4b}
\hat{G}=\sum_{\lambda_\nu}
\hat{R}^{-1}\frac{-\lambda_{\nu}\hat{1}}{\omega^2+\lambda_{\nu}^2}\hat{R}\;.
\end{eqnarray}

The correlation function between the currents $I_{\alpha}$ and $I_{\alpha'}$
can be written in the form~\cite{korotkov}
\begin{equation}\label{b5}
S_{\alpha\alpha'}(\omega)=\delta_{\alpha,\alpha'}
S^{Sch}_{\alpha}+S^c_{\alpha\alpha'}(\omega)\;,
\end{equation}
where
\begin{equation}\label{b6}
S^{Sch}_{\alpha}\equiv 2e(I^+_{\alpha}+I^-_{\alpha})=
2e^2\sum_{n<m}(\Gamma^{nm}_{\alpha} p_{m}+\Gamma^{mn}_{\alpha} p_{n})\;
\end{equation}
is the high frequency ($\omega\to \infty$) limit of the shot-noise (the
Schottky noise). The frequency-dependent part is expressed as~\cite{korotkov}
\begin{eqnarray}\label{b7}
S^c_{\alpha\alpha'}(\omega)=\pm 2e^2\{ \sum_{n,m}
[\Gamma^{mm+1}_{\alpha}-\Gamma^{mm-1}_{\alpha}]G_{mn}(\omega)
[\Gamma^{n-1n}_{\alpha'}p_{n-1}-\Gamma^{n+1n}_{\alpha'}p_{n+1}]\nonumber\\ +
\sum_{n,m}[\Gamma^{mm+1}_{\alpha'}-\Gamma^{mm-1}_{\alpha'}]G_{mn}(-\omega)
[\Gamma^{n-1n}_{\alpha}p_{n-1}-\Gamma^{n+1n}_{\alpha}p_{n+1}]\} \;,
\end{eqnarray}
where the sign ($+$) is for the cross-correlation function between the currents
in the source and the drain electrode, the sign ($-$) is for the case with both
currents in the drain electrodes or in the source electrode.

All results for the currents and the correlation functions can be derived
analytically using a symbolic mathematical program. A problem is presentation
of the results, because the formulae can be very complex. Below,
we present the results for few interesting cases from a physical point of
view.

\subsection{Medium voltage range - Two levels in the voltage window}

We consider the case of a moderate source-drain voltage $V$, for which two
energy levels $E_1$ and $E_2$ lie in the voltage window, but the level $E_3$ is
beyond the voltage window range and it does not participate in transport.
Electrons can be transferred only from the left to the right hand side in
Fig.\ref{fig1} and backflow is ignored. Nonzero elements for the total
tunneling rates are: $\Gamma^{01}_L=\gamma^{01}_L$,
$\Gamma^{02}_L=\gamma^{02}_L$, $\Gamma^{10}_{R_1}=\gamma^{10}_{R_1}$,
$\Gamma^{20}_{R_1}=\gamma^{20}_{R_1}$, $\Gamma^{10}_{R_2}=\gamma^{10}_{R_2}$,
$\Gamma^{20}_{R_2}=\gamma^{20}_{R_2}$. We ignore smearing of the Fermi
distribution function $f^{\pm}_{\alpha,n}$
 and they are taken equal to 1 or 0, respectively, in the expressions for the
total tunneling rates $\Gamma^{nm}_{\alpha}$. In this voltage regime the
Schottky term $S^{Sch}_{\alpha}$ [Eq.(\ref{b6})] is equal to the Poissonian
value. However, for lower voltages or higher temperatures, when backflow is
allowed, $S^{Sch}_{\alpha}$ increases and can lead to the super-Poissonian shot
noise.\cite{bb99,bb00,safonov} In this paper we focus only on dynamical
processes, which can lead to an enhancement of shot noise seen in the term
$S^c_{\alpha\alpha'}(\omega)$ [Eq.(\ref{b7})]. Therefore, our studies are
restricted to the case $h \gamma^{nm}_{\alpha}\ll k_BT\ll eV$. In this regime
one can find analytical solutions and can easily extract various processes
contributing to shot noise.

Let us begin with a simplest case, when the transfer rates to/from the states $E_1$ and
$E_2$ are equal,  i.e. $\gamma^{01}_L=\gamma^{02}_L=\gamma_{L}$,
$\gamma^{10}_{R_1}=\gamma^{20}_{R_1}=\gamma_{R_1}$,
$\gamma^{10}_{R_2}=\gamma^{20}_{R_2}=\gamma_{R_2}$.  The current and the
correlation functions at $\omega=0$ can be expressed as
\begin{eqnarray}\label{b8}
I_{R_1}=\frac{2e \gamma_L\gamma_{R_1}}{2\gamma_L + \gamma_{R_1} + \gamma_{R_2}}\;,\\
\label{b9}
S_{R_1,R_1}(0)=2e|I_{R_1}|\;\left[1-\frac{4\gamma_L\gamma_{R_1}}{(2\gamma_L +
\gamma_{R_1}
+ \gamma_{R_2})^2}\right]\;,\\
\label{b10}
S_{R_1,R_2}(0)=\frac{-16e^2\gamma_L^2\gamma_{R_1}\gamma_{R_2}}{(2\gamma_L +
\gamma_{R_1}
 + \gamma_{R_2})^3}\;.
\end{eqnarray}
As one could expect anti-bunching processes dominate in this system.
The auto-correlation function $S_{R_1,R_1}(0)$, for the
currents in the drain electrodes, is always reduced, and the cross-correlation
function $S_{R_1,R_2}(0)$ is always negative.

Now we consider an ideal splitter,  i.e. the case when an electron can be
transferred only from the level $E_1$ to the $R_1$ electrode, but transfer from
this level to the $R_2$ electrode is forbidden, and similarly, for transfers
from the level $E_2$, which can be realized to the $R_2$ electrode only. Thus,
the nonzero tunneling rates are: $\gamma^{10}_{R_1}=\gamma_{R_1}$,
$\gamma^{20}_{R_2}=\gamma_{R_2}$ and the transfer rates from the source
electrode are still kept the same $\gamma^{01}_{L}=\gamma^{02}_{L}=\gamma_L$.
The current, the auto- and the cross-correlation function are expressed as
\begin{eqnarray}\label{b11}
I_{R_1}=I_{R_2}=\frac{e
\gamma_{L}\gamma_{R_1}\gamma_{R_2}}{\gamma_{L}\gamma_{R_1} +
\gamma_{L}\gamma_{R_2} + \gamma_{R_1}\gamma_{R_2}}\;,\\
\label{b12}
 S_{R_1,R_1}(0)=2e|I_{R1}|\left\{1-\frac{2
\gamma_L\gamma_{R_1}[\gamma_{R_2}(\gamma_L +\gamma_{R_2}) -
\gamma_L\gamma_{R_1}]}{(\gamma_{L}\gamma_{R_1} +
\gamma_{L}\gamma_{R_2} + \gamma_{R_1}\gamma_{R_2})^2}\right\}\;,\\
\label{b13}
S_{R_1,R_2}(0)=\frac{-2e^2\gamma_{L}^2\gamma_{R_1}\gamma_{R_2}[\gamma_{R_1}
\gamma_{R_2} (\gamma_{R_1} +\gamma_{R_2})
-\gamma_L(\gamma_{R_1}-\gamma_{R_2})^2] }{(\gamma_{L}\gamma_{R_1} +
\gamma_{L}\gamma_{R_2} + \gamma_{R_1}\gamma_{R_2})^3}\;.
\end{eqnarray}
At $\omega=0$ the current conservation is fulfilled and one can derive all
other correlation functions using Eq.(\ref{b11})-(\ref{b13}), e.g.
$S_{R_1L}(0)=S_{R_1R_1}(0)+S_{R_1R_2}(0)$.

Although the tunneling rates $\gamma_{R_1}$ and $\gamma_{R_2}$ are different,
the currents (\ref{b11}) in both the drain electrodes are the same (as it
should be for an ideal splitter). The effect is due to dynamical Coulomb
blockade, the process which distributes electrons to both the electrodes with
the same probability. From Eq.(\ref{b12}) one can determine a condition
$\gamma_L(\gamma_{R_1}-\gamma_{R_2})>\gamma_{R_2}^2$ for the super-Poissonian
shot noise in the $R_1$ electrode. Positive cross-correlation between the
currents in the drain electrodes $R_1$ and $R_2$ are for
$\gamma_L(\gamma_{R_1}-\gamma_{R_2})^2>\gamma_{R_1}\gamma_{R_2}(\gamma_{R_1}+\gamma_{R_2})$.
Both conditions are fulfilled for large asymmetry between the outgoing
channels, i.e. for tunneling rates $\gamma_{R_1}\gg \gamma_{R_2}$.

\begin{figure}[ht]
\centerline{\epsfxsize=0.35\textwidth \epsfbox{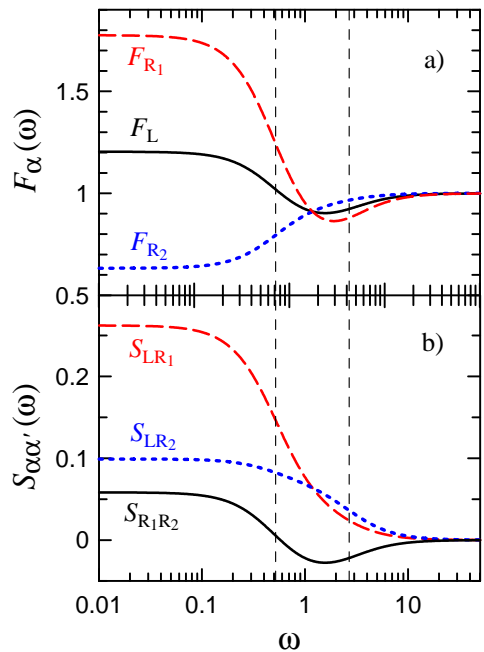} }
\caption{\label{fig2}(Color online) Frequency dependence of the Fano factors
$F_{L}$, $F_{R_1}$, $F_{R_2}$ (black, red, blue curves) - Fig.a, and the
current cross-correlation functions $S_{R_1R_2}$, $S_{LR_1}$, $S_{LR_2}$
(black, red, blue curves) - Fig.b, for a medium voltage window in the
Y-terminal system with an ideal two level splitter [for the transfer rates
$\gamma_{L}=5$, $\gamma_{R_1}=1$, $\gamma_{R_2}=0.5$, for which the inverse of
the relaxation time is $1/\tau=0.74$ and 10.75 (vertical dashed lines)]. The
electron charge is taken as $e=-1$ in all our presentations. }
\end{figure}

In order to have an insight into dynamical processes leading to
super-Poissonian shot noise and positive cross correlations we perform a
frequency analysis of the auto- and cross-correlation functions. Fig.\ref{fig2}
presents the results for a large asymmetry between the tunneling rates in the
outgoing channels. The zero-frequency of the Fano factors $F_{R_1}(0)$,
$F_L(0)>1$ and the cross-correlations $S_{R_1R_2}(0)$, $S_{LR_1}(0)$,
$S_{LR_2}(0)$ are positive. It is clearly seen that all correlation functions
have two contributions corresponding to two dynamical processes with
characteristic frequencies $\lambda_1=0.74$ and $\lambda_2=10.75$. For most
of the correlation functions the low frequency process leads to an enhancement of the
auto- and the cross-correlation function, whereas the high frequency process
gives a negative contribution. An exception is $S_{L,R_2}(\omega)$, for which
the lower frequency component is negative and the high frequency contribution
is positive. It means that various dynamical processes can occur in a system
with electron-electron interactions, which can lead to bunching or anti-bunching. The
bunching process is generally for lower frequency fluctuations, but in some
cases higher frequency fluctuations can lead to bunching as well.

\subsection{Three levels in the voltage window}\label{sec-three}

Now we consider a large voltage regime, for which  three energy levels $E_1$,
$E_2$ and that one $E_3=E_1+E_2+U_{12}$ for two electrons in QD,  are in the
voltage window and all of them participate in transport. The results are
presented for an ideal splitter, i.e. when an electron can be only transferred
from $E_1$ to the $R_1$ electrode, from $E_2$ to the $R_2$ electrode and
transfer from the $L$ electrode to all three levels is with the same tunneling
rate. For this case the nonzero tunneling rates are:
$\gamma^{10}_{R_1}=\gamma^{32}_{R_1}=\gamma_{R_1}$,
$\gamma^{20}_{R_2}=\gamma^{31}_{R_2}=\gamma_{R_2}$ and
$\gamma^{01}_L=\gamma^{02}_L=\gamma^{13}_L=\gamma^{23}_L=\gamma_L$.
 The current and the correlation functions are
expressed by
\begin{eqnarray}\label{b14}
I_{R_1}=\frac{e \gamma_{L}\gamma_{R_1}}{\gamma_{L}+\gamma_{R_1}}\;,\\
\label{b15} S_{R_1,R_1}(\omega)=2e|I_{R_1}|\left[1-\frac{2
\gamma_{L}\gamma_{R_1}}
 {\omega^2+(\gamma_{L}+\gamma_{R_1})^2}\right]\;,\\
 \label{b16}
S_{R_1,R_2}(\omega)=0\;.
\end{eqnarray}
The current $I_{R_2}$ can be derived from (\ref{b14}) exchanging $\gamma_{R_1}$
and $\gamma_{R_2}$, and $I_L=I_{R_1}+I_{R_2}$. Eq.(\ref{b14})
suggests that $I_{R_1}$ is independent of the tunneling rate $\gamma_{R_2}$ and
of the current $I_{R_2}$ in the second drain electrode. Indeed both the
currents are independent, which one can see from Eq.(\ref{b16}) for the cross-correlation
function $S_{R_1,R_2}(\omega)=0$. These formulae (\ref{b14})-(\ref{b16}) are much
simpler than those (\ref{b11})-(\ref{b13}) for the two levels in the voltage window (when
double occupancy of the splitter is forbidden). From comparison of these two cases
and one see a role of strong correlations in
bunching of electrons and enhancement of the auto- and the cross-correlation
functions.

Naively one could expect that shot noise should increase for the splitter with
an increase of a number of levels, due to an increase of fluctuations through
new tunneling channels. A spectral decomposition analysis helps us to understand
a relation between fluctuations and their
contribution to the correlations functions. Using (\ref{b4b}) for the Green
function we can express
\begin{eqnarray}\label{b15a}
 S_{R_1,R_1}(\omega)=2e|I_{R_1}|\left[1-\sum_{\lambda_{\nu}} \frac{s_{R_1,R_1}(\lambda_{\nu})}
 {\omega^2+\lambda_{\nu}^2}\right]\;,
\end{eqnarray}
where the eigenfrequencies are: $\lambda_1=\gamma_L+\gamma_{R_1}$,
$\lambda_2=\gamma_L+\gamma_{R_2}$, and
$\lambda_3=2\gamma_L+\gamma_{R_1}+\gamma_{R_2}$. The coefficients are
$s_{R_1,R_1}(\lambda_{1})=2\gamma_{L}\gamma_{R_1}$, and
$s_{R_1,R_1}(\lambda_{2})=s_{R_1,R_1}(\lambda_{3})=0$. We separate tunneling
processes through the level $E_1$ and $E_3$ in the current
$I_{R_1}(t)=I_{R_1}^{(1)}(t)+I_{R_1}^{(3)}(t)$.
 The coefficients are written in the form
\begin{eqnarray}\label{b15b}
s_{R_1,R_1}(\lambda_{\nu}) =s^{(11)}_{R_1,R_1}(\lambda_{\nu})+2
s^{(13)}_{R_1,R_1}(\lambda_{\nu})+ s^{(33)}_{R_1,R_1}(\lambda_{\nu})\;,
\end{eqnarray}
where $s^{(nm)}_{R_1,R_1}(\lambda_{\nu})$ are the coefficients corresponding to the
current-current correlation functions through the levels $n,m=1,3$. Derivations
are not complex, and we get $s^{(11)}_{R_1,R_1}(\lambda_{1})=\gamma_{R_1}
\gamma_{R_2}^2\gamma_L/(\gamma_{R_2} +\gamma_L)^2$,
$s^{(11)}_{R_1,R_1}(\lambda_{2})= -\gamma_{R_1}
\gamma_{R_2}\gamma_L^2/[(\gamma_{R_1}+\gamma_L)
 (\gamma_{R_2}+\gamma_L)]$, $s^{(11)}_{R_1,R_1}(\lambda_{3})= \gamma_{R_1} \gamma_{R_2} \gamma_L^2
(\gamma_{R_1}+\gamma_{R_2}+2 \gamma_L)/ [(\gamma_{R_1}+\gamma_L)
(\gamma_{R_2}+\gamma_L)^2]$, $s^{(13)}_{R_1,R_1}(\lambda_{1})=\gamma_{R_1}
\gamma_{R_2} \gamma_L^2/ (\gamma_{R_2}+\gamma_L)^2$, $
s^{(33)}_{R_1,R_1}(\lambda_{1})=\gamma_{R_1}
\gamma_L^3/(\gamma_{R_2}+\gamma_L)^2$. Moreover,  the coefficients
$s^{(11)}_{R_1,R_1}(\lambda_{2})=-s^{(13)}_{R_1,R_1}(\lambda_{2})=
s^{(33)}_{R_1,R_1}(\lambda_{2})$ and
$s^{(11)}_{R_1,R_1}(\lambda_{3})=-s^{(13)}_{R_1,R_1}(\lambda_{3})=
s^{(33)}_{R_1,R_1}(\lambda_{3})$ for the eigenfrequencies
$\lambda_2$ and $\lambda_3$, respectively. This analysis shows that for $\lambda_2$ and
$\lambda_3$ the inter-level cross-correlation functions have opposite sign to
the auto-correlation parts and they compensate each other. The nonzero
contribution to the shot noise (\ref{b15}) is from the components
$s^{(11)}_{R_1,R_1}(\lambda_{1})$, $s^{(13)}_{R_1,R_1}(\lambda_{1})$ and
$s^{(33)}_{R_1,R_1}(\lambda_{1})$, only. A similar analysis of the
cross-correlation function $S_{R_1,R_2}(\omega)$ shows total compensation of
various partial correlation functions for the currents through all three levels.

\subsection{Different potentials in drain electrodes}

\begin{figure}[ht]
\centerline{\epsfxsize=0.3\textwidth \epsfbox{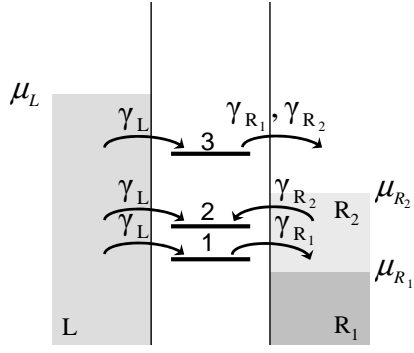} } \caption{\label{fig3}
Schematic presentation of tunneling for the Y-terminal system with a
three-level splitter and with different chemical potentials $\mu_{\alpha}$ in
the $R_1$ and $R_2$ electrode. The tunneling rates, which participate in
transport are presented.}
\end{figure}

In the Y-terminal system one can apply different voltages to the drain
electrodes and get two different voltage windows for electron transport.
Fig.{\ref{fig3} presents a situation with a large voltage window for the $R_1$
electrode and a medium voltage window for the second drain electrode. Thus, an
electron can tunnel from QD to the $R_1$ electrode from $E_3$ and $E_1$.
Tunneling to the $R_2$ electrode can be only from $E_3$ (only when two
electrons occupy QD); tunneling from the single electron state $E_2$ is
forbidden because it lies below $\mu_{R_2}$. The tunneling rates are assumed to
be the same as for an ideal splitter, for which the nonzero tunneling rates
are: $\gamma^{01}_L=\gamma^{02}_L=\gamma^{13}_L=\gamma^{23}_L=\gamma_L$,
$\gamma^{10}_{R_1}=\gamma^{32}_{R_1}=\gamma_{R_1}$,
$\gamma^{02}_{R_2}=\gamma^{31}_{R_2}=\gamma_{R_2}$. The currents and the
cross-correlation functions are expressed by
\begin{eqnarray}\label{b17}
I_{R_1}=\frac{e \gamma_{L}\gamma_{R_1}}{\gamma_{L}+\gamma_{R_1}}\;,\\\label{b18}
I_{R_2}=\frac{e\gamma_{L}^2\gamma_{R_2}}
{(\gamma_{L}+\gamma_{R_1})(\gamma_{L}+\gamma_{R_2})}\;,\\\label{b19}
S_{R_1,R_2}(0)=-\frac{2e^2\gamma_L^2\gamma_{R_1}\gamma_{R_2}(\gamma_L-\gamma_{R_1})}
{(\gamma_L+\gamma_{R_1})^3(\gamma_L+\gamma_{R_2})}\;,\\\label{b20}
S_{R_1,R_1}(0)=2e|I_{R_1}|\left[1-\frac{2 \gamma_L\gamma_{R_1}}
 {(\gamma_{L}+\gamma_{R_1})^2}\right]\;,\\\label{b21}
S_{R_2,R_2}(0)=2e|I_{R_2}|\Bigl\{1-2 \gamma_{R_2}\bigl[\gamma_{R_1}^3(\gamma_L+\gamma_{R_2})^2+
 \gamma_L^4(\gamma_{R_2}+2\gamma_L)
 +\gamma_L^3\gamma_{R_1}(\gamma_{R_2}+3\gamma_L)\nonumber\\+
\gamma_{R_1}^2\gamma_L(2\gamma_{R_2}^2+4\gamma_{R_2}\gamma_L+
 3\gamma_L^2)\bigr]
/\bigl[\gamma_L(\gamma_{L}+\gamma_{R_1})^2(\gamma_{L}+\gamma_{R_2})^2(2\gamma_L+\gamma_{R_1}
+\gamma_{R_2})\bigr]\Bigr\}\;.\end{eqnarray}
The current $I_{R_1}$ and the
auto-correlation function $S_{R_1,R_1}(0)$ [Eq.(\ref{b17}) and (\ref{b19})] are
the same as for the case of three levels in the voltage window
[Eq.(\ref{b14})-(\ref{b15})] considered in the previous section. The
cross-correlation function $S_{R_1,R_2}(0)$ is positive for
$\gamma_{R_1}>\gamma_L$. Performing spectral decomposition (\ref{b4b}) one can
see contribution of relaxation processes with characteristic eigenfrequencies:
$\lambda_1=\gamma_L+\gamma_{R_1}$, $\lambda_2=\gamma_L+\gamma_{R_2}$,
$\lambda_3=2\gamma_L+\gamma_{R_1}+\gamma_{R_2}$. The frequency-dependent
correlation functions are
\begin{eqnarray}\label{b23}
S_{R_1,R_2}(\omega)=\frac{2e^2\gamma_L\gamma_{R_1}\gamma_{R_2}}
{(\gamma_{R_1}-\gamma_{R_2})(\gamma_{R_1}+\gamma_{R_2})(2\gamma_L+\gamma_{R_1}+\gamma_{R_2})}
\nonumber\\ \times \left[
\frac{\gamma_{R_1}(\gamma_{R_1}^2-\gamma_{R_1}\gamma_{R_2}+\gamma_L\gamma_{R_1}-
2\gamma_L^2-3\gamma_L\gamma_{R_2})} {\omega^2+(\gamma_L+\gamma_{R_1})^2}+
\frac{\gamma_{R_2}(\gamma_{L}+\gamma_{R_2})(2\gamma_L-\gamma_{R_1}+\gamma_{R_2})}
{\omega^2+(\gamma_L+\gamma_{R_2})^2}
\right]\;,\\
S_{R_1,R_1}(\omega)=2e|I_{R_1}|\left[1-\frac{2 \gamma_{L}\gamma_{R_1}}{\omega^2+
(\gamma_L+\gamma_{R_1})^2}\right]\;,\\
S_{R_2,R_2}(\omega)=2e|I_{R_2}|\biggl\{1-\frac{2\gamma_{R_2}}
{\gamma_L(\gamma_{R_1}-\gamma_{R_2})
(2\gamma_L+\gamma_{R_1}+\gamma_{R_2})}\nonumber\\\times
\biggl[\frac{\gamma_{R_1}^2(\gamma_L+\gamma_{R2})
(2\gamma_L+\gamma_{R_1})}{\omega^2+(\gamma_L+\gamma_{R_1})^2}\nonumber\\
-\frac{\gamma_{R_1}^2 (\gamma_L+\gamma_{R_2})^2 -
\gamma_L^2\gamma_{R_2}(2\gamma_L+\gamma_{R_2}) +
\gamma_L\gamma_{R_1}(\gamma_{R_2}^2 + 2\gamma_L\gamma_{R_2} + 2
\gamma_L^2)}{\omega^2+(\gamma_L+\gamma_{R_2})^2}\biggr]\biggr\}\;.
\end{eqnarray}
One can see that fluctuations corresponding to the eigen-frequencies
$\lambda_1$ and $\lambda_2$ only contribute  to the current correlation
functions. A fluctuation process with the characteristic frequency
$\lambda_3$ is absent in these correlation functions.

\section{Y-terminal system with a quantum dot splitter}

\subsection{Mediate voltage range}

\begin{figure}[ht]
\centerline{\epsfxsize=0.4\textwidth \epsfbox{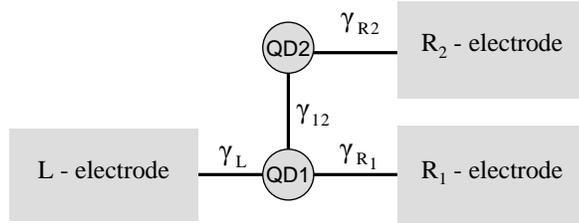} }
\caption{\label{figa1} Schematic presentation of the Y-terminal system with two
quantum dots. The net transfer rates between the quantum dots and the
electrodes are shown.}
\end{figure}

 Now, we consider the dynamical Coulomb blockade process in a three-terminal system
 connected by two quantum dots (presented
schematically in Fig.\ref{figa1}). The system is similar to two capacitatively
coupled quantum dots~\cite{michalek} and an experimental setup~\cite{zhang}
(see the second part of the paper [\onlinecite{zhang}] with a Y-terminal
structure), where the sub and super-Poissonian shot noise were studied.
Similarly as in the previous section, we consider sequential electronic
transport, which is governed by the classical master
equation~\cite{korotkov,gurvitz96,gurvitz98}
\begin{eqnarray}\label{1}
\frac{d}{dt} \left[\begin{array}{c} p_2\\p_1\\p_0\end{array}\right]= {\hat
M}\;\left[\begin{array}{c} p_2\\p_1\\p_0\end{array}\right],
\end{eqnarray}where
\begin{eqnarray}\label{1a}
\hat M = \left[\begin{array}{ccccc}
-\gamma_{R_2}&\gamma_{12}&0\\
0&-\gamma_{12}-\gamma_{R_1}&\gamma_{L}\\
\gamma_{R_2}&\gamma_{R_1}&0&-\gamma_{L}
\end{array}\right]\;.
\end{eqnarray}
Here, $p_0$, $p_1$ and $p_2$ denote probabilities finding an empty system, a system with
one electron either at the quantum dot (QD1) or (QD2). The applied voltage
difference to the left and the right electrodes is assumed to be moderate, and
therefore, electron transfers are only from left to right and given by the net
transfer rates shown in Fig.\ref{figa1}. Moreover, it is assumed that only one
electron can be at the splitter, either at QD1, or QD2, and $p_2+p_1+p_0=1$.

The currents flowing into the right electrodes are given by
\begin{eqnarray}\label{3}
 I_{R_1}=e\gamma_{R_1}p_1=\frac{e\gamma_{L}\gamma_{R_1}
\gamma_{R_2}}{w}\;,\\\label{5} I_{R_2}=e\gamma_{R_2}p_2=\frac{e\gamma_{L}
\gamma_{12}\gamma_{R_2}}{w}\;,
\end{eqnarray}
where $w=\gamma_{L}\gamma_{12} +\gamma_{L}\gamma_{R_2}+\gamma_{12}
\gamma_{R_2}+\gamma_{R_1} \gamma_{R_2}$. For $\gamma_{12}=\gamma_{R_1}$ the
current $I_{R_1}=I_{R_2}$ and the system of two QD acts as an ideal splitter
(for any value of $\gamma_{R_2}$ and $\gamma_{L}$).

The current-current correlation functions
can be written as~\cite{korotkov}
\begin{equation}\label{6}
S_{\alpha,\alpha'}(\omega)=\delta_{\alpha,\alpha'}
2eI_{\alpha}+S^c_{\alpha,\alpha'}(\omega)\;,
\end{equation}
where the frequency dependent parts are
\begin{eqnarray}\label{8}
S^c_{LL}(\omega)=4e^2\;\gamma_{L}\mathrm{Re}[G_{01}(\omega)]\gamma_{L}p_0\;,\\
\label{9}
S^c_{R_1R_1}(\omega)=4e^2\;\gamma_{R_1}\mathrm{Re}[G_{10}(\omega)]\gamma_{R_1}p_1\;,\\
\label{10}
S^c_{R_2R_2}(\omega)=4e^2\;\gamma_{R_2}\mathrm{Re}[G_{20}(\omega)]\gamma_{R_2}p_2
\end{eqnarray}
for the auto-correlation functions and
\begin{eqnarray}\label{11}
S_{LR_1}(\omega)=2e^2\left\{\gamma_{L}\mathrm{Re}[G_{00}(\omega)]\gamma_{R_1}p_1+
\gamma_{R_1}\mathrm{Re}[G_{11}(\omega)]\gamma_{L}p_0\right\}\;,\\
\label{12}
S_{LR_2}(\omega)=2e^2\left\{\gamma_{L}\mathrm{Re}[G_{00}(\omega)]\gamma_{R_2}p_2+
\gamma_{R_2}\mathrm{Re}[G_{21}(\omega)]\gamma_{L}p_0\right\}\;,\\
\label{13}
S_{R_1R_2}(\omega)=2e^2\left\{\gamma_{R_1}\mathrm{Re}[G_{10}(\omega)]\gamma_{R_2}p_2
+ \gamma_{R_2}\mathrm{Re}[G_{20}(\omega)]\gamma_{R_1}p_1\right\}\;.
\end{eqnarray}
for the cross-correlation functions. From Eqs.(\ref{8})-(\ref{10}) one can see
that the autocorrelation functions $S_{\alpha\alpha}$  are proportional to the
current $I_{\alpha}$ [Eq.(\ref{3})-(\ref{5})]. The  factor
$F_{\alpha}=S_{\alpha\alpha}/2eI_{\alpha}$ is known as the Fano factor. The
cross-correlation functions $S_{\alpha\alpha'}$ [Eq.(\ref{11})-(\ref{13})] have
two terms proportional to the current $I_{\alpha}$ and $I_{\alpha'}$,
respectively. One can not extract a single factor proportional neither to the
current $I_{\alpha}$, nor $I_{\alpha'}$, nor $\sqrt{I_{\alpha}I_{\alpha'}}$. It
is not possible to define the Fano factor for the cross-correlation functions
[see also Eq.(\ref{17}) for $S_{R_1R_2}$ presented below]. In the literature
~\cite{cottetprb,sanchez,vaseghi} different normalizations of the
cross-correlation noise were used and they were called the Fano factor, but in
our opinion they have no physical meaning.

At the zero frequency limit
the Fano factors for the auto-correlation functions are given by
\begin{eqnarray}\label{14}
F_{L}=1+2\gamma_{L}^2\gamma_{12}\gamma_{R_1}/w^2 -2 \gamma_{L} \gamma_{R_2}
\bigl[\gamma_{12}(\gamma_{L}+\gamma_{12}+\gamma_{R_1}+
\gamma_{R_2})+\gamma_{R_1}\gamma_{R_2}
\bigr]/w^2 \;,\\
\label{15}
F_{R_1}=1+2\gamma_{L}\gamma_{R_1}(\gamma_{L}\gamma_{12}-\gamma_{R_2}^2)/w^2\;,\\
\label{16} F_{R_2}=1-2\gamma_{L}\gamma_{12}
\gamma_{R_2}(\gamma_{L}+\gamma_{12}+\gamma_{R_1}+\gamma_{R_2})/w^2\;.
\end{eqnarray}
In these formulae (\ref{14})-(\ref{16}) we extracted positive and negative
terms responsible for the super-Poissonian and the sub-Poissonian shot noise.
From Eq.(\ref{15}) one can easily see that $F_{R_1}$ is in the super-Poissonian
regime, if the transfer rate $\gamma_{R_2}$ is small, and
when one can expect a large charge accumulation at QD2. Fig.\ref{fig5}a
presents that the Fano factor $F_L$ is also larger than unity at small
$\gamma_{R_2}$. The factor $F_{R_2}< 1$ for any transfer rates. It is clear
that dynamical Coulomb blockade does not occur in the $R_2$ channel and shot noise
is always the sub-Poissonian type. The reduction of $F_{R_2}$ can be
substantial, and it can drop to the minimal value $F_{R_2}=1/3$ for
$\gamma_{L}=\gamma_{12}=\gamma_{R_2}$ and $\gamma_{R_1}\to 0$ [see
Eq.(\ref{16})]. It is well known that the Fano factor in sequential transport
through a quantum dot may be reduced to its minimal value
$F=1/2$.~\cite{korotkov92,hershfield,korotkov,blanter} In the present case,
however,  the system is with two quantum dots connected in series. In the
Appendix we show that the shot noise reduction can be even larger for a system
with a large number of quantum dots.
The exact calculations show that in a system of $N$ quantum dots connected in
series the Fano factor can be reduced to the value $F=1/(N+1)$.

\begin{figure}[ht]
\centerline{\epsfxsize=0.35\textwidth \epsfbox{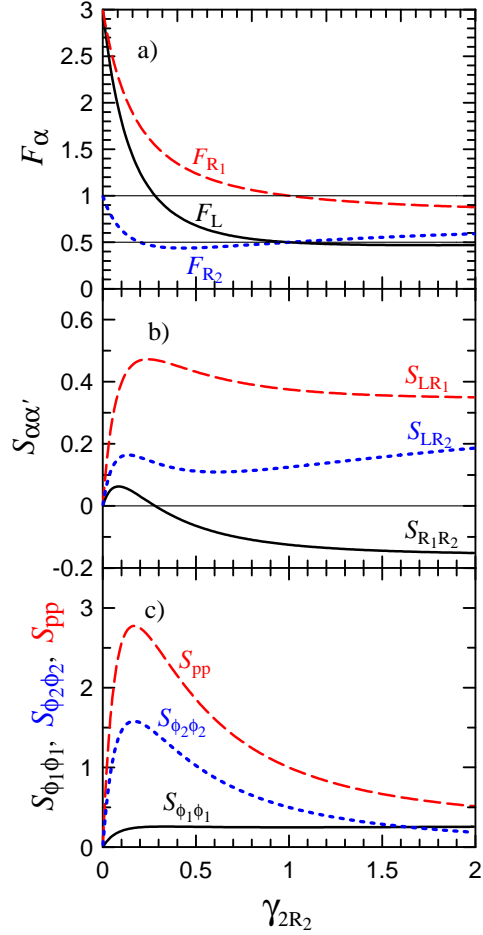}}
 \caption{\label{fig5}
(Color online) Plot of the Fano factors $F_{\alpha}$ (for the $L$-, $R_1$ and
$R_2$ electrodes - black, red and blue curve, respectively) -- Fig.a; the
current cross-correlation functions $S_{\alpha\alpha'}$ ($S_{R_1R_2}$ - black,
$S_{LR_1}$ - red and $S_{LR_2}$ - blue) -- Fig.b; and the potential correlation
functions $S_{\phi_1\phi_1}$, $S_{\phi_2\phi_2}$ (black and blue) as well as
the polarization correlation function $S_{pp}$ (red) -- Fig.c, as a function of
the transfer rate $\gamma_{R_2}$ at the zero-frequency $\omega=0$. The other
transfer rates are: $\gamma_{L}=\gamma_{12}=\gamma_{R_1}=1$. }
\end{figure}

Using the formulae (\ref{11})-(\ref{13}) one can derive the cross-correlations functions.
Below we present the function
\begin{eqnarray}\label{17}
S_{R_1R_2}(0)=-2I_{R_1}I_{R_2}(\gamma_{L}+\gamma_{12}+\gamma_{R_1}+2
\gamma_{R_2})/w +2e^2\gamma_{L}^3\gamma_{12}^2\gamma_{R_1}\gamma_{R_2}/w^3\;.
\end{eqnarray}
At $\omega=0$ the current conservation is fulfilled and we can derive all other
cross-correlation functions using Eq.(\ref{14})-(\ref{17}), e.g.
$S_{LR_1}(0)=S_{R_1R_1}(0)+S_{R_2R_1}(0)$,
$S_{LR_2}(0)=S_{R_1R_2}(0)+S_{R_2R_2}(0)$. Fig.\ref{fig5}b presents the results
for the cross-correlation functions. All of them show a peak at a small $\gamma_{R_2}$.
Increasing the tunneling rate $\gamma_{R_2}$ the correlation function $S_{R_1R_2}$
decreases and changes its sign. It suggests that bunching processes weaken their intensity
and antibunching becomes dominating.

Bunching effects and
super-Poissonian noise are caused by dynamical Coulomb blockade, which should be seen in
charge and potential fluctuations in the system (especially at QD2). If we denote a local potential
$\phi_i(t)=\phi_0+en_i(t)$ depending on fluctuation of the number of electrons
$n_i(t)$ at the i-th quantum dot, then using Eq.(\ref{b4}) one can write the
potential-potential correlation function
\begin{eqnarray}\label{18}
S_{\phi_i\phi_i}(0)\equiv 4\sum_{n,m}\phi_i(n)\mathrm{Re}[G_{nm}(0)]\phi_i(m)p_{m}
=4e^2\mathrm{Re}[G_{ii}(0)]p_i\;.
\end{eqnarray}
Fig.\ref{fig5}c shows that the potential fluctuations $S_{\phi_2\phi_2}$ at QD2
 are large in the region of small values of $\gamma_{R_2}$. In
Fig.\ref{fig5}c we plotted the correlation function $S_{pp}$ for the
polarization $p(t)=\phi_2(t)-\phi_1(t)$ between the charges localized at QD1
and QD2. The corresponding curve (red) presents very large fluctuations
in the same region of the parameters as the super-Poissonian shot noise.
Note that $S_{\phi_1\phi_1}$  monotonically increases and it does not show
activation of any potential fluctuations  at small $\gamma_{R_2}$.

\begin{figure}[ht]
\centerline{\epsfxsize=0.4\textwidth \epsfbox{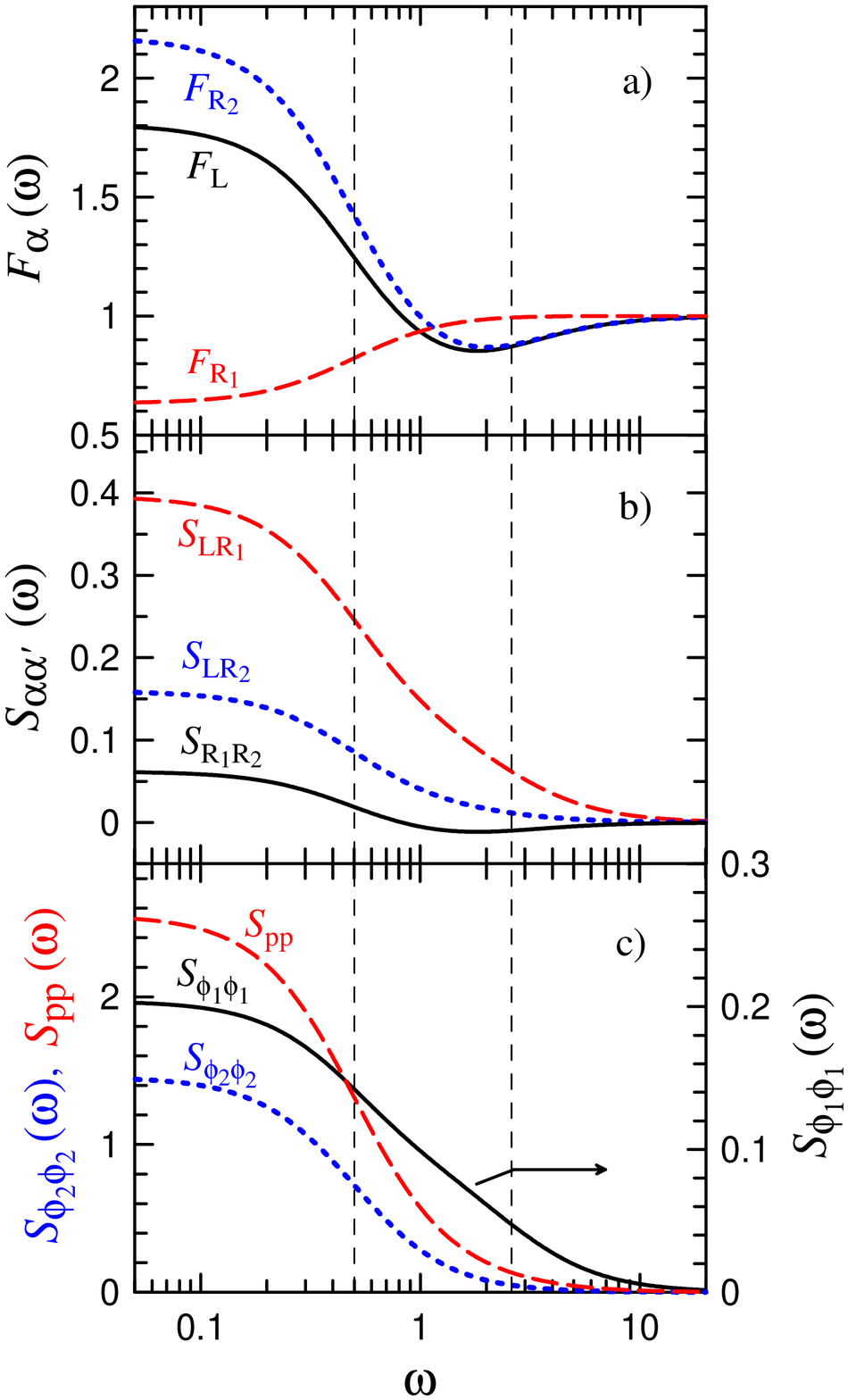}} \caption{\label{fig6}
(Color online) Frequency dependence of the Fano factors $F_{L}$, $F_{R_1}$,
$F_{R_2}$ (black, red, blue curves) - Fig.a; the current cross-correlation
functions $S_{R_1R_2}$, $S_{LR_1}$, $S_{LR_2}$ (black, red, blue curves) -
Fig.b; and the potential correlation functions $S_{\phi_1\phi_1}$,
$S_{\phi_2\phi_21}$  and the polarization correlation function $S_{pp}$ (black,
blue and red curves) - Fig.c. The transfer rates are:
$\gamma_{L}=\gamma_{12}=\gamma_{R_1}=1$, $\gamma_{R_2}=0.1$, for which the
eigen-frequency is $\lambda=0.5$ and 2.6 (vertical dashed lines).}
\end{figure}

These results suggest that a charge accumulated at QD2 influences the
transport through QD1. In order to get more information on dynamics
of this feedback process
 we perform a frequency analysis. One can do a spectral decomposition of
all studied correlation functions and determine components corresponding to
eigen-frequencies of the system (see eg.[\onlinecite{bb00}]). For the case
presented in Fig.\ref{fig6} one gets $\lambda_1=0.5$ and $\lambda_2=2.6$. Its
inverse $\tau_{\nu}=1/\lambda_{\nu}$ is a relaxation time for
eigen-fluctuations in the system. A role of the corresponding relaxation
processes in the range of the super-Poissonian noise is presented in
Fig.\ref{fig6}. The Fano factors $F_{L}$ and $F_{R_2}$ are higher than unity in
the low frequency range, whereas $F_{R_1}$ is always below unity. It means that
for $F_{L}$ and $F_{R_2}$ their low frequency components lead to
super-Poissonian shot noise, whereas high frequency contributions reduce shot
noise. For given parameters the low frequency part can dominate the high
frequency contribution and a measurement should show a super-Poissonian value
of a zero-frequency power spectrum. Fig.\ref{fig6}b shows frequency-dependent
plots of the cross-correlation functions. The function  $S_{R_1R_2}$ clearly
shows two components: low and high frequency ones. Low frequency components are
positive and they dominate. It means that the low frequency process leads to
the positive cross-correlation function $S_{R_1R_2}$ for outgoing electrons.

 Fig.\ref{fig6}c
presents fluctuations of the polarization $S_{pp}$ and the charge fluctuations
$S_{\phi_2\phi_2}$ at QD2, which occur only in the low frequency
regime. The plot of the function $S_{\phi_1\phi_1}$ is different and shows that
the high frequency component is also relevant in the potential fluctuations at
QD1. Applying the spectral decomposition procedure we can assign the
lower eigen-frequency to the polarization fluctuations. It means also that the
dynamical Coulomb blockade leads to the bunching effect for electrons.

\subsection{Large voltage window}

In Sec.\ref{sec-three}, we showed that if two electron states participate in
transport through a multi-level splitter, then the dynamical Coulomb blockade effect is reduced, and in a special
case the cross-correlation function can be $S_{R_1R_1}(\omega)=0$. One can expect also
a reduction of shot noise for the quantum dot splitter at higher voltages. In
this section we consider a situation with a large voltage window when energy of
second electron overcomes the Coulomb
repulsion energy and the electron is introduced on an empty QD. The master equation is
\begin{eqnarray}\label{45}
\frac{d}{dt} \left[\begin{array}{c} p_{12}\\p_2\\p_1\\p_0\end{array}\right]=
{\hat M}\;\left[\begin{array}{c} p_{12}\\p_2\\p_1\\p_0\end{array}\right],
\end{eqnarray}where
\begin{eqnarray}\label{46}
\hat M = \left[\begin{array}{cccc}
-\gamma_{R_1}-\gamma_{R_2}&\gamma_{L}&0&0\\
\gamma_{R_1}&-\gamma_{L}-\gamma_{R_2}&\gamma_{12}&0\\
\gamma_{R_2}&0&-\gamma_{12}-\gamma_{R_1}&\gamma_{L}\\
0&\gamma_{R_2}&\gamma_{R_1}&-\gamma_{L}
\end{array}\right]\;,
\end{eqnarray}
and $p_{12}$ is the probability to find simultaneously two electrons at QD1 and
QD2. The currents flowing into the right electrodes are
\begin{eqnarray}\label{47}
I_{R_1}=\frac{e\gamma_{L}\gamma_{R_1}(\gamma_{R_1}\gamma_{R_2}+\gamma_{R_2}^2
+\gamma_{L}\gamma_{12}+\gamma_{L}\gamma_{R_2})}{w}\;,\\\label{48}
I_{R_2}=\frac{e\gamma_{L}\gamma_{12}\gamma_{R_2}(\gamma_{L}+\gamma_{R_1}+\gamma_{R_2})}{w}\;,
\end{eqnarray}
where $w=\gamma_{R_2} (\gamma_{R_1}+\gamma_{L})
(\gamma_{R_1}+\gamma_{R_2}+\gamma_{L})+\gamma_{12}
[\gamma_{R_2}^2+\gamma_{R_2} \gamma_{L}+\gamma_{L}^2+\gamma_{R_1}
(\gamma_{R_2}+\gamma_{L})]$. When the two electron state $(1,1)$ participates
in transport the currents increase [compare Eq.(\ref{47})-(\ref{48}) with
Eq.(\ref{3})-(\ref{5})], but their enhancement is different
for the different drain electrodes.
In previous section we have seen that the system acts as an ideal current
splitter (with $I_{R_1}=I_{R_2}$) for $\gamma_{12}=\gamma_{R_1}$.
Now, because single and two
electron states are in the voltage window, ideal current redistribution
is broken (in general, $I_{R_1}\neq I_{R_2}$).

\begin{figure}[ht]
\centerline{\epsfxsize=0.35\textwidth \epsfbox{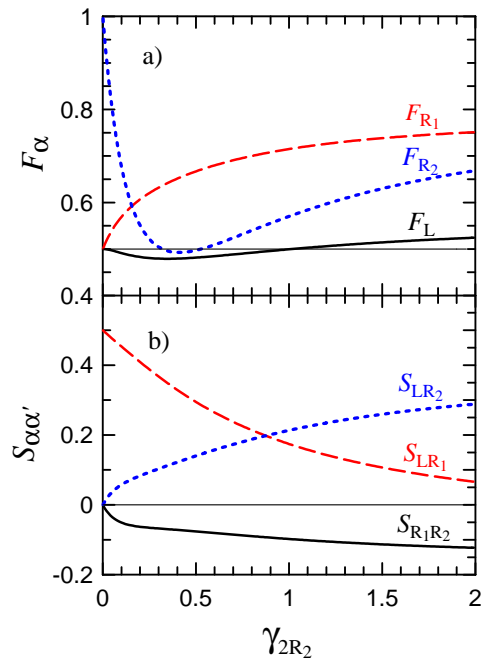}} \caption{\label{fig7}
(Color online) Plot of the Fano factors $F_{\alpha}$ (for the $L$-, $R_1$ and
$R_2$ electrodes - black, red and blue curve, respectively) -- Fig.a, and the
current cross-correlation functions $S_{\alpha\alpha'}$ ($S_{R_1R_2}$ - black,
$S_{LR_1}$ - red and $S_{LR_2}$ - blue) -- Fig.b, as a function of the transfer
rate $\gamma_{R_2}$ at $\omega=0$ and in the high voltage regime, when single
and two electron states participate in transport. The parameters are the same
as in Fig.5.}
\end{figure}

 The current-current correlation functions are derived analytically in the same way as
 previously. Because the formulae are rather complex, we present plots of the auto- and
 cross-correlation functions in Fig.\ref{fig7}. All Fano factors are in the
 sub-Poissonian regime. The factors $F_{L}$ and $F_{R_2}$ show strong reduction, which
 can be even below the value $F=1/2$ for a single quantum dot system. The
 cross-correlation function $S_{R_1R_2}$ is always negative. These results are different
 than those presented in Fig.\ref{fig5} for a mediate voltage regime, when the two electron
 state is outside the voltage window. Now the correlation functions do not show
 bunching effects -- antibunching processes dominate in shot noise.

\section{Summary and concluding remarks}

Our study addresses dynamical aspects in current shot noise in the Y-terminal
system with a multilevel splitter.  We showed that the zero-frequency
current-current auto- and cross-correlation functions increase because of the
dynamical Coulomb blockade effect. Furthermore, the spectral decomposition
analysis indicates that the dynamic Coulomb blockade effect is caused by charge
fluctuations corresponding to an electron accumulated at the level with a
lowest outgoing tunneling rate. This process exhibits itself a large asymmetry
in the tunneling rates and strong Coulomb interactions. In the section II.B, we
present the case for the Y-terminal system with an ideal two-level splitter,
when an electron is transferred from a given level to only one of two drain
electrodes, and with a further assumption of a single electron occupancy of the
splitter. An increase of the voltage changes the situation dramatically. For a
large voltage, when three levels are in the voltage window and double electron
occupancy is allowed, electrons transferred to the drain electrodes behave like
independent particles with the cross-correlation function
$S_{R_1,R_2}(\omega)=0$ (for any frequency $\omega$). Having separated the
currents into the partial currents, flowing through each energy level, and
using the spectral analysis, we were able to decompose the correlation
functions, and determine a role of each individual component for any
eigenfrequency. The partial auto-correlation functions compensate the partial
cross-correlation functions, leading to $S_{R_1,R_2}(\omega)=0$, and reduce the
auto-correlation function $S_{R_1,R_1}(\omega)$ [see Eq.(\ref{b17})]. For the
situation considered, bunching and antibunching scattering processes compensate
each other perfectly. We also considered the case with different voltages
applied to the drain electrons, with the voltage windows of different sizes
(with three and one level in the window, respectively). The current
fluctuations are then compensated only partially. For example, the
cross-correlations can be positive and negative, depending on the tunneling
rates for incoming and outgoing channels.

We suggest to perform an experiment to see these correlation effects on a
multilevel QD system. A direct observation of the bunching process on a
two-terminal multilevel QD was performed by Gustavsson et
al.~\cite{gustavssonprl,gustavssonprb} using a quantum point contact system
acting as a charge sensor. Our proposal is different because the system should
be three-terminal one with asymmetric tunnel contacts, and current cross
correlation measurements should be performed. In the experiment, one has to
change the voltage window, from the low to the strong voltage regime, in order
to change the number of levels and the number of electrons participating in
transport.

We also considered the Y-terminal system with two quantum dots, which is the
model corresponding to the experimental setup recently studied by Zhang et
al.~\cite{zhang}. In the experiment~\cite{zhang}, one can control all the
parameters of the system, and verify our theoretical predictions. Our studies
show that the dynamical Coulomb blockade effect leads to electron bunching, and
the effect manifests itself with strong charge fluctuations at the side quantum
dot (QD2).  In order to activate the effect, one has to apply the gate voltage,
and reduce the tunneling rate $\gamma_{R_2}$ from QD2 to the electrode. We
predict an increase of the Fano factor $F_{R_1}$ to the super-Poissonian regime
and positive cross-correlations $S_{R_1R_2}$ for $\gamma_{R_2}\to 0$. When the
bias voltage increases, we expect that the two electron state will become a
participant in transport, and shot noise will significantly be reduced to the
sub-Poissonian regime.  We also predict that in a system consisting of N
quantum dots connected in series, a minimal value of the Fano factor is
$F=1/(N+1)$.

\begin{figure}[ht]
\centerline{\epsfxsize=0.35\textwidth \epsfbox{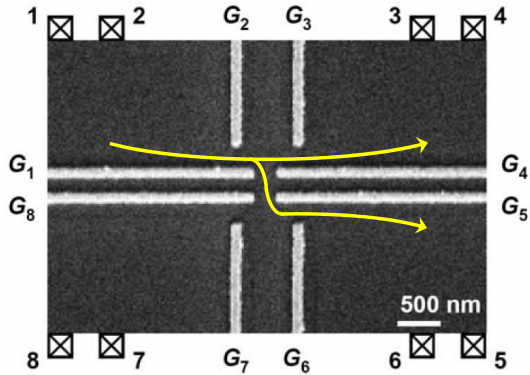}} \caption{\label{fig8}
An experimental setup for two coupled quantum wires, which can be used for
studies of shot noise and electron bunching (adapted from [\onlinecite{bird}]).
By tuning voltages applied to the gates G$2$, G$_3$ and G$_6$, one can change tunneling
transfer rates for the incoming and outgoing channels in the splitter, and test
the dynamical Coulomb blockade processes in the current-current cross
correlations.}
\end{figure}

We could not point out a microscopic origin of an enhancement of shot noise in
the experiments on a quantum point contact~\cite{chen,zarchin}. We also could
not find a source of local or spatial charge (potential) fluctuations, which
lead to bunching in a such simple experimental setup with a quantum point
contact. We hope that dynamical Coulomb blockade is again responsible for
bunching in this case. Therefore, we propose shot noise measurements in a
similar system presented in Fig.\ref{fig8}. In this case, two ballistic
channels are coupled with each other, and current can be split into two
electrodes, similar to the experiments in [\onlinecite{chen, zarchin}]. This
experimental setup allows control of output channels by the gate electrodes
G$_3$ and G$_6$, which can also control charge fluctuations in the system. We
speculate that dynamical Coulomb blockade leads to bunching in coherent
electron transport as well. To our best knowledge, there is no research in the
coherent transport regime showing electron bunching. While there are related
papers~\cite{noisekondo} on shot noise in the coherent regime, which take into
account electron-electron interactions and the Kondo resonance, but their shot
noise is in the sub-Poissonian regime.

\acknowledgments{I would like to acknowledge stimulating  discussions with Jerzy Wr{\'o}bel.
The work was supported as part of the European Science
Foundation EUROCORES Programme FoNE by funds from the Ministry of Science and
Higher Education and EC 6FP (contract N. ERAS-CT-2003-980409), and the EC
project RTNNANO (contract N. MRTN-CT-2003-504574).}

\appendix*
\section{Current noise in quantum dots connected in series}

We consider a system of $N$-quantum dots connected in series in a high voltage limit,
in which electrons can only hop from left to right. The master equation can be written as
\begin{eqnarray}\label{a1}
\frac{d}{dt} \left[\begin{array}{c} p_N\\\vdots\\p_2\\p_1\\p_0\end{array}\right]=
\left[\begin{array}{cccccc}
-\gamma_{NR}&\gamma_{N-1N}&...&0&0&0\\
\vdots&\vdots&\ddots&\vdots&\vdots&\vdots\\
0&0&...&-\gamma_{23}&\gamma_{12}&0\\
0&0&...&0&-\gamma_{12}&\gamma_{L1}\\\
\gamma_{NR}&0&...&0&0&-\gamma_{L1}\\
\end{array}\right]\;\left[\begin{array}{c}  p_N\\\vdots\\p_2\\p_1\\p_0\end{array}\right],
\end{eqnarray}
where we have assumed a single electron in the system, i.e. $\sum_n p_n=1$. Here,
we present the analytical results for the $N=3$ system. The current is expressed as
\begin{eqnarray}\label{a2}
I=e\frac{\gamma_{L1}\gamma_{12}\gamma_{23}\gamma_{3R}}{\gamma_{L1}\gamma_{12}\gamma_{23}
+\gamma_{L1}\gamma_{12}\gamma_{3R}+\gamma_{L1}\gamma_{23}\gamma_{3R}+
\gamma_{12}\gamma_{23}\gamma_{3R}}\;.
\end{eqnarray}
The zero-frequency current shot noise is given by
\begin{eqnarray}\label{a3}
S_{LL}=S_{RR}=S_{LR}=2eI\frac{\gamma_{L1}^2\gamma_{12}^2\gamma_{23}^2
+\gamma_{L1}^2
\gamma_{12}^2\gamma_{3R}^2+\gamma_{L1}^2\gamma_{23}^2\gamma_{3R}^2+
\gamma_{12}^2\gamma_{23}^2\gamma_{3R}^2}{(\gamma_{L1}\gamma_{12}\gamma_{23}
+\gamma_{L1}\gamma_{12}\gamma_{3R}+\gamma_{L1}\gamma_{23}\gamma_{3R}+
\gamma_{12}\gamma_{23}\gamma_{3R})^2}\;.
\end{eqnarray}
It is clear that the Fano factor $F=S/2eI\leq 1$ and its lowest value $\min\{F\}=1/4$
is reached for $\gamma_{L1}=\gamma_{12}=\gamma_{23}=\gamma_{3R}$.
Generalization for any $N$ can
be straightforward performed by a mathematical induction, showing that
$\min\{F\}=1/(N+1)$.

This result is a generalization of the Fano factor derived for a single QD in a
sequential tunneling regime, when
$\min\{F\}=1/2$.~\cite{korotkov92,hershfield,korotkov,blanter}

\end{document}